\documentclass[%
 reprint,
superscriptaddress,
 amsmath,amssymb,
 aps,
pra,
]{revtex4-2}

\usepackage[pdftex]{graphicx}
\usepackage{dcolumn}
\usepackage{physics}
\usepackage{bm}
\usepackage{xcolor}
\usepackage{tikz}

\providecommand{\V}[1]{\boldsymbol{#1}}
\providecommand{\M}[1]{\mathbf{#1}}
\providecommand{\T}[1]{\mathrm{#1}}

\providecommand{\J}{\T{i}}

\begin{document}





\preprint{APS/123-QED}

\title{Characteristic Mode Analysis of Acoustic Scatterers}

\author{Lukas Jelinek}
\affiliation{Department of Electromagnetic Field, Czech Technical University in Prague}
\email{lukas.jelinek@fel.cvut.cz}

\author{Kurt Schab}
\affiliation{School of Engineering, Santa Clara University}

\author{Viktor Hruska}
\affiliation{Department of Physics, Czech Technical University in Prague}

\author{Miloslav Capek}
\affiliation{Department of Electromagnetic Field, Czech Technical University in Prague}

\author{Mats Gustafsson}
\affiliation{Department of Electrical and Information Technology, Lund University}

\date{\today}

\begin{abstract}
The explicit connection between the transition matrix and boundary element method integral operators is formulated.  This enables the calculation of characteristic modes via eigenvalue problems involving either set of operators, leading to convenient orthogonality properties facilitating scattering analysis, solution of inverse problems, and the design of excitation fields.
\end{abstract}

\maketitle


\section{Introduction}

Modal decomposition is an important part of mathematical physics when studying linear wave phenomena~\cite{MorseFeshbachMethodsOfTheoreticalPhysics}. It is especially common in quantum mechanics, electromagnetics, optics, acoustics, or structural dynamics, where complex vibrations are decomposed into uncoupled constituents, allowing insight into the underlying dynamics. Typically, the so-called normal modes~\cite{AryaIntroductionToClassicalMechanics} are studied, being eigenstates of the differential equation governing the physical system (most commonly the Helmholtz equation). This leads to modes of optical waveguides~\cite{CollinFieldTheoryOfGuidedWaves}, modes of elastic membranes~\cite{KinslerFundamentalsOfAcoustics}, or stationary states of quantum objects~\cite{CohenTannoudjiPhotonsAndAtoms}, to name a few examples.

The eigenstates of integral operators~\cite{ColtonIntegralEquationMethodsInScatteringTheory} are much less commonly studied, despite having favorable properties when employed in scattering problems. An important exception is the use of eigenmodes of the scattering operator in inverse acoustic scattering~\cite{mast1997focusing} or using characteristic modes to analyze and design radiating electromagnetic structures~\cite{CMA_review2022_Part1_Lau}.

In this paper, we adapt the concept of \textit{characteristic modes} (as they are defined and used throughout electromagnetics) to the study of acoustic scatterers.  Along the way, we discuss their properties, and, importantly, derive their relation to the acoustic boundary element method (BEM)~\footnote{In this paper we will generally use BEM to denote formulation of wave scattering via integral equations and subsequent use of Galerkin's method to transform them into linear equation system.}, a vital tool for analyzing acoustic scattering~\cite{KirkupTheBoundaryElementMethodInAcoustics}. As part of this analysis, we also discuss the explicit link between scattering/transition matrices and BEM integral operators, allowing for fast numerical computation of the former from the latter.

\section{Characteristic Modes}

Characteristic modes form a specific basis resulting from the eigenvalue decomposition of integral operators related to scattering problems.  The particular form of this decomposition leads to favorable modal properties, such as orthogonality of scattering patterns and quantification of scattering strength, making them widely used in electromagnetic theory and design (see~\cite{CMA_review2022_Part1_Lau,CMA_review2022_Part2_Capek,CMA_review2022_Part3_Adams,CMA_review2022_Part4_Li,CMA_review2022_Part5_Manteuffel} for extensive references).  This section introduces the notation and properties of characteristic modes as they are used in electromagnetics~\cite{harrington1971theory}. Nevertheless, the notation and approach described here are quite general and applicable to any linear wave phenomena.  Specific details of its application and interpretation in acoustics problems are the focus of all subsequent sections.

\begin{figure}
    \centering
    \includegraphics[width=3.25in]{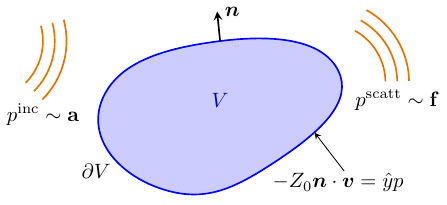}
    \caption{Scattering problem setup where an incident pressure $p^\T{inc}$ impinges on a volume $V$ bounded by an impedance boundary with normalized surface admittance $\hat{y}$.  The induced total surface pressure $p$ and normal velocity $\V{n} \cdot \V{v}$ produce the scattered pressure $p^\T{scatt}$.}
    \label{fig:bem-schem}
\end{figure}

In a time-harmonic steady state at angular frequency~$\omega$~\footnote{A time convention~$\T{exp} \left(- \J \omega t \right)$ with $\J = \sqrt{-1}$ denoting the imaginary unit is assumed. Notice that as compared to the works in electrical engineering journals, this leads to sign changes at several places, notably of characteristic number~$\lambda_n$.}, the characteristic modes of an object are defined as the eigenmodes~\cite{garbacz1965modal} of its transition matrix
\begin{equation}
    \M{T}\M{a}_n = t_n\M{a}_n,
    \label{eq:t-eig}
\end{equation}
see Fig.~\ref{fig:bem-schem}.  The transition matrix $\M{T}$ describes the scattering nature of the obstacle by mapping incident pressure coefficients~$\M{a}$ to scattered pressure coefficients~$\M{f}$ (in some appropriate basis, discussed in later sections)
via~\cite{waterman1969new,Mishchenko_ComprehensiveList2016}
\begin{equation}
    \M{f} = \M{T}\M{a}.
    \label{eq:fta}
\end{equation}
Hence, the eigenvectors $\M{a}_n$ obtained in \eqref{eq:t-eig} describe characteristic excitations (incident pressure distributions) expressed in the selected basis.  In electromagnetics~\cite{garbacz1965modal,harrington1971theory}, the complex eigenvalue~$t_n$ is commonly represented by a characteristic number~$\lambda_n$ via relation~\cite{garbacz1965modal,Gustafsson_etal2021_CMAT_Part1}
\begin{equation}
    t_n = -\dfrac{1}{1- \J\lambda_n}\quad\leftrightarrow\quad \lambda_n = - \J \left( 1 + t_n^{-1} \right) .
    \label{eq:tn-lambdan}
\end{equation}
In the study of perfectly conducting scatterers, the occurrence of a zero characteristic number $\lambda_n = 0$ is associated with resonance (balance between cycle-mean electric and magnetic stored energy) and is often a desired design objective.  From \eqref{eq:tn-lambdan}, this condition coincides with a maximum magnitude of the eigenvalue $|t_n| = 1$, indicating a maximal interaction of the scatterer with the characteristic excitation described by $\M{a}_n$.

\begin{figure}
    \centering
    \includegraphics[width=3.25in]{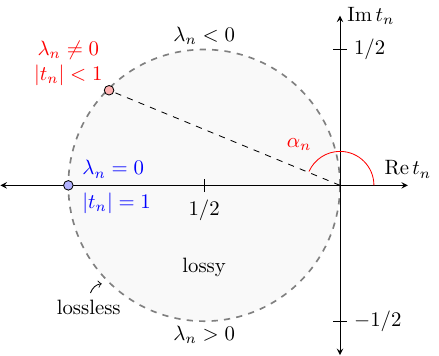}
    \caption{Schematic depiction of the relationships between characteristic mode eigenvalue $t_n$, characteristic number $\lambda_n$, and characteristic angle $\alpha_n$.}
    \label{fig:eigenvalue-representations}
\end{figure}

For lossless systems~\cite{Gustafsson_etal2021_CMAT_Part1,Gustafsson_etal2021_CMAT_Part2}, the values~$t_n$ lie on a circle in the complex plane, see Fig.~\ref{fig:eigenvalue-representations}.  For lossy (dissipative) systems, the eigenvalues lie within this circle. In analogy to electromagnetic systems, the sign of real-valued number~$\lambda_n$ (for lossless systems) can also be attributed to the excess of cycle-mean potential or kinetic energy (electric or magnetic energy in electromagnetics). In other words, the sign of~$\lambda_n$ coincides with the sign of cycle-mean Lagrangian density~\cite[Chap. 6.2]{MorseTheoreticalAcoustics}.  Further, characteristic number~$\lambda_n$ can be represented by a characteristic angle
\begin{equation}
    \alpha_n = \angle t_n = \pi + \arctan \lambda_n
\end{equation}
with $\alpha_n = \pi$ indicating resonance.  This angle is limited to the range $\pi/2 \leq \alpha_n \leq 3\pi/2$.

Eigenvectors~$\M{a}_n$ represent the expansion coefficients of the modal fields in the selected basis used to represent the transition operator as the matrix $\M{T}$, such as spherical vector waves (electromagnetics)~\cite{Kristensson_ScatteringBook}, or scalar spherical waves (acoustics)~\cite{MorseTheoreticalAcoustics}. For lossless scatterers, the eigenvectors~$\M{a}_n$ are orthogonal, since the transition matrix is normal~\cite{2012_WeiJiang2012_TUFF,Gustafsson_etal2021_CMAT_Part1}. This implies that characteristic modes of a lossless scatterer exhibit orthogonal far-field patterns. Furthermore, the eigenvectors can be made real-valued for lossless reciprocal scenarios~\cite{Gustafsson_etal2021_CMAT_Part1}. Other characteristic quantities, such as the characteristic surface pressure or velocity, can be computed by exciting the system with the vector~$\M{a}_n$ and evaluating the desired quantity~\cite{2012_WeiJiang2012_TUFF,Gustafsson_etal2021_CMAT_Part1}.

These orthogonality properties of characteristic modes and the close link to the integral equations~\cite{harrington1971computation,capek2022computational} make them particularly attractive for many design and analysis applications in electromagnetics.  In particular, diagonalization of the system matrix (often associated with an impedance-like operator) describing the scattering scenario leads to simple formulas for modal excitation coefficients in terms of incident fields~\cite[Eq.~30]{harrington1971theory}, which in turn aids in the design of antenna elements or the placement of small antennas on larger platforms, see \cite{CMA_review2022_Part3_Adams} and \cite{CMA_review2022_Part4_Li} for extended bibliographies on these topics.

The frequency dependence of the integral operators used in forming the characteristic mode eigenvalue problem means all modal quantities are inherently functions of frequency.  Since these quantities can only be computed at finite numbers of frequencies, a procedure connecting and interpolating continuous ``tracked'' modal data across frequency is required.  The problem of modal tracking has been studied extensively in the electromagnetics community~\cite{CMA_review2022_Part2_Capek} with solutions including numerical correlation~\cite{RainesRojas_WidebandTrackingOfCharModes}, analytic symmetry-based rules \cite{SchabBernhard_GroupTheoryForCMA,Maseketal_ModalTrackingBasedOnGroupTheory}, and methods utilizing special properties of scattering operators~\cite{Gustafsson_etal2021_CMAT_Part2}. When frequency sweeps of characteristic data are presented in this paper, tracking based on the correlation of scattered far fields at neighboring frequencies~\cite[\S-II]{Gustafsson_etal2021_CMAT_Part2} is employed.
  
  Computation-wise, characteristic modes in electromagnetic scattering are often approached using Galerkin's formulation of field integral equations~\cite{harrington1971theory,harrington1971computation,capek2022computational}. Such connection in acoustics is one of the main contributions of this paper and is detailed in the following sections.



\section{Boundary Element Method}
\label{sec:cm-bem}

Assume a time-harmonic steady state and consider a closed surface in Fig.~\ref{fig:bem-schem} with outer unit normal~$\V{n}$ which is homogeneously covered by a surface impedance~$Z_\T{s}$ relating total surface pressure~$p$ and its normal derivative by
\begin{equation}
     \J \dfrac{1}{k_0}\V{n} \cdot \nabla p = \hat{y} p \Leftrightarrow  - Z_0 \V{n} \cdot \V{v} = \hat{y} p.
     \label{eq:surfImpedCond}
\end{equation}
Here~$\hat{y}$ is normalized surface admittance $\hat{y} = Z_0/Z_\T{s}$,~$k_0 = \omega/c_0$ is background wavenumber, and~$Z_0 = \rho_0 c_0$ is background impedance with~$\rho_0$ and $c_0$ being background density and speed of sound, respectively~\cite{MorseTheoreticalAcoustics,KinslerFundamentalsOfAcoustics}. We also note that, in source-free regions, the velocity field is $\V{v} = \left( \J k_0 Z_0 \right)^{-1} \nabla p$.  If the surface pressure is induced by an incident wave~$p^\T{inc}$ produced by sources external to the closed surface, the scattered pressure is given~\cite[Chap.~8]{MorseTheoreticalAcoustics} by the addition of two components~$p^\T{scatt}_p \left\{ p \right\}$, $p^\T{scatt}_{\T{d}p} \left\{ \V{n} \cdot \nabla p  \right\}$ associated with the surface pressure and its normal derivative.  The exact forms of these fields are detailed in Appendix~\ref{app:A}. Employing the boundary condition~\eqref{eq:surfImpedCond}, the integral equation relating incident and total surface pressure can be formed as~\cite{1986_Amini_CMAME,bendali2008boundary}
\begin{equation}
    \left(\mathcal{D} - \J k_0 \hat{y} \mathcal{S}\right)\left\{ p \right\} = - p^\T{inc},
    \label{eq:general-bem}
\end{equation}
where
\begin{equation}
    \mathcal{S} \left\{ p \right\} = (-\T{i}k_0\hat{y})^{-1}p^\T{scatt}_{\T{d}p} \left\{\V{n}\cdot\nabla p  \right\} = p^\T{scatt}_{\T{d}p} \left\{p  \right\}
\end{equation}
and
\begin{equation}
    \mathcal{D} \left\{ p \right\} = - p + p^\T{scatt}_p \left\{ p \right\}
\end{equation} 
are related to the single- and double-layer potential operators~\cite{bendali2008boundary}, respectively.  Note that the operator $\mathcal{S}$ is symmetric, while $\mathcal{D}$ is not.  This has implications on modal orthogonality properties, discussed in subsequent sections.

Expanding the surface pressure into basis functions~$\psi_n$
\begin{equation}
    p(\V{r}) = \sum_n p_n\psi_n(\V{r}),\quad \M{p} = 
    \mqty[p_1 & p_2& ...& p_N]^\T{T},
\end{equation}
with~$^\T{T}$ denoting transposition,
and employing Galerkin's method~\cite{Kantorovich1982},~\cite[\S 2]{kaltenbacher2018computational} allows the transformation of the scattering operators into matrices so that equation~\eqref{eq:general-bem} is written as a system of linear equations 
\begin{equation}
    \left(\M{D} - \J k_0 \hat{y} \M{S} \right)\M{p} = - \M{p}^\T{inc},
    \label{eq:general-bemGalerkin}
\end{equation}
see Appendix~\ref{app:A} for details.  By the definition of the surface impedance \eqref{eq:surfImpedCond}, the normal derivative of pressure $\V{n}\cdot\nabla p$ or normal velocity $\V{n}\cdot \V{v}$ can be expanded in the same basis $\{\psi_n\}$ with coefficients $- \J k_0\hat{y}p_n$.

The connection of acoustic BEM~\eqref{eq:general-bemGalerkin} and the characteristic mode eigenvalue problem in \eqref{eq:t-eig} is best shown by expanding the total pressure field external to the scatterer into scalar spherical waves~\cite[Sec.~3.5,~9.6]{Jackson_ClassicalElectrodynamics} 
\begin{equation}
    p\left( \V{r} \right) = k_0 \sqrt{Z_0} \sum\limits_\beta  \left( a_\beta \T{u}_\beta ^{\left( 1 \right)} \left( k_0\V{r} \right) + f_\beta \T{u}_\beta^{\left( 3 \right)} \left( k_0\V{r} \right) \right),
    \label{eq:swExp}
\end{equation}
where coefficients~$a_\beta$ represent the incident field (functions $\T{u}_\beta ^{\left( 1 \right)}$ contain spherical Bessel functions), while coefficients~$f_\beta$ represent the scattered field (functions $\T{u}_\beta ^{\left( 3 \right)}$ contain outgoing spherical Hankel functions). The normalization is made such that the cycle-mean scattered power is given by, see Appendix~\ref{app:B},
\begin{equation}
    P^\T{scatt} = \dfrac{1}{2} \sum_\beta \left| f_\beta \right|^2.
\end{equation}

The next step is to connect the spherical expansion vectors~$\M{a}, \M{f}$ with the expansion vector of the surface pressure~$\M{p}$. This is initiated through the decomposition of the background Green's function into scalar spherical waves~\cite[Sec.~9.6]{Jackson_ClassicalElectrodynamics} and leads to a set of matrices~$ \M{U}, \M{U}^p, \M{U}^{\T{d}p}$, detailed in Appendix~\ref{app:A}, that provide this connection. Namely, the expansion vector~$\M{f}$ of outgoing spherical waves generated by the surface pressure~$\M{p}$ is given by
\begin{equation}
    \M{f} = - \M{U}\M{p} = - (\M{U}^p - \J k_0 \hat{y}\M{U}^{\T{d}p})\M{p},
    \label{eq:UmatDef}
\end{equation}
and, similarly, the incident pressure expansion vector~$\M{p}^\T{inc}$ can be obtained from spherical expansion vector~$\M{a}$ as
\begin{equation}
    \M{p}^\T{inc} = \J k_0 Z_0 \M{U}^{\T{d}p,\T{H}} \M{a},
    \label{eq:pIncExp}
\end{equation}
where $^\T{H}$ denotes Hermitian (conjugate) transpose.

Inverting the system matrix in~\eqref{eq:general-bemGalerkin} and using~\eqref{eq:UmatDef} and~\eqref{eq:pIncExp} directly leads to the system's transition matrix 
\begin{equation}
    \M{T} = \J k_0 Z_0\M{U}\left(\M{D} - \J k_0 \hat{y} \M{S}\right)^{-1}\M{U}^{\T{d}p,\T{H}}.
    \label{eq:t-mat}
\end{equation}
Formula~\eqref{eq:t-mat} presents the connection between characteristic modes~\eqref{eq:t-eig} and BEM~\eqref{eq:general-bemGalerkin} and closely resembles similar links established in computational electromagnetics between impedance-like operators and transition matrices~\cite{gurel1991connection,Gustafsson_etal2021_CMAT_Part1}. In the following section, we discuss the implications of this relation on the computation of characteristic modes. We also note that formula~\eqref{eq:t-mat} presents a way to obtain the transition matrix of a scatterer of arbitrary complexity and parallels similar formulas in electromagnetic theory~\cite{2013_Kim_TAP, 2017_Markkanen_JQSRT, 2022_Losenicky_TAP}.

\section{Characteristic Modes Evaluated from BEM matrices}
\label{sec:CMfromBEM}

Similarly to electromagnetic problems, acoustic problems admit evaluation of characteristic modes either in terms of their transition matrix~$\M{T}$ or directly from BEM matrices.  In this section, we outline this equivalence for general surface impedances and study the limiting cases associated with hard or pressure-release boundaries.


Particularly, inserting the transition matrix~\eqref{eq:t-mat} into eigenvalue problem~\eqref{eq:t-eig} gives
\begin{equation}
    \J k_0 Z_0\M{U}\left(\M{D} - \J k_0 \hat{y} \M{S}\right)^{-1}\M{U}^{\T{d}p,\T{H}} \M{a}_n = t_n\M{a}_n.
\end{equation}
Left multiplying with the matrix $\M{U}^{\T{d}p,\T{H}}$ and small rearrangements lead to the eigenvalue problem
\begin{equation}
    \J \left(\M{D} - \J k_0 \hat{y} \M{S}\right) \M{p}_n = \left( 1- \J\lambda_n \right) k_0 Z_0 \M{U}^{\T{d}p,\T{H}} \M{U} \M{p}_n
    \label{eq:BEMcm1}
\end{equation}
with eigenvectors
\begin{equation}
    \M{p}_n = - \J k_0 Z_0 \left(\M{D} - \J k_0 \hat{y} \M{S}\right)^{-1}\M{U}^{\T{d}p,\T{H}} \M{a}_n
\end{equation}
serving as the characteristic surface pressure, see~\eqref{eq:general-bem} and~\eqref{eq:pIncExp}. Relation~\eqref{eq:BEMcm1} may be thought of as another way to evaluate characteristic modes, in this case, directly from BEM matrices $\M{D}$ and $\M{S}$ and the transformation matrix $\M{U}$.

When the system is lossless,~$\T{Re} \left\{ \hat{y} \right\} = 0$, further simplification is possible. For that, factorization of the Green's function into spherical waves is used to derive
\begin{equation}
    \T{Re}\left\{ \J \M{D}\right\} = k_0 Z_0\M{U}^{\T{d}p,\T{H}}\M{U}^p
    \label{eq:power-balance-1}
\end{equation}
and
\begin{equation}
    \T{Re}\left\{ \J \M{S}\right\} = k_0 Z_0\M{U}^{\T{d}p,\T{H}}\M{U}^{\T{d}p},
    \label{eq:power-balance-2}
\end{equation}
where, in contrast to decomposition into Hermitian and anti-Hermitian parts~\cite{HornJohnson_MatrixAnalysis}, here $\T{Re}\left\{\M{A}\right\}$ (and later $\T{Im}\left\{\M{A}\right\}$)  denotes the element-wise real (imaginary) part of a matrix $\M{A}$. Relations~\eqref{eq:power-balance-1} and~\eqref{eq:power-balance-2} are employed (again relying on the assumption of a lossless admittance~$\hat{y}$) to write
\begin{equation}
    k_0 Z_0 \M{U}^{\T{d}p,\T{H}} \M{U} = 
\T{Re}\left\{  \J \left(\M{D} - \J k_0 \hat{y} \M{S}\right) \right\},
\end{equation}
which after substituting into~\eqref{eq:BEMcm1}, gives
\begin{equation}
    \M{X} \M{p}_n = -\lambda_n \M{R} \M{p}_n 
    \label{eq:xr}
\end{equation}
with 
\begin{subequations}
\begin{equation}
    \M{X} = \T{Im} \left\{ \J \left(\M{D} - \J k_0 \hat{y} \M{S}\right) \right\}, 
\end{equation}
\begin{equation}
    \M{R} = \T{Re} \left\{\J \left(  \M{D} - \J k_0 \hat{y} \M{S}  \right) \right\}.
\end{equation}
\end{subequations}
This is the equation for characteristic modes based solely on BEM matrices without the need of the transition matrix $\M{T}$ or mapping matrices~$\M{U}$. We note that eigenvalues~$\lambda_n$ are real-valued under the assumption of a lossless system.  However, compared to the classical treatment in electromagnetics with electric field integral equation~\cite{harrington1971theory}, the characteristic pressures cannot be made equiphase since the matrices~$\M{R}$ and~$\M{X}$ are not symmetric, except in the particular case of pressure release boundaries ($\hat{y} = \pm\T{i}\infty$) when only the symmetric matrix $\M{S}$ contributes to these matrices.

\section{Expansion of Driven Problems Into Characteristic Modes}

For a lossless system, the transition matrix $\M{T}$ is normal and the characteristic excitations~$\M{a}_n$ are orthogonal
\begin{equation}
    \M{a}_m^\T{T}\M{a}_n = \M{a}_n^\T{T}\M{a}_n \delta_{mn},
    \label{eq:a-orth}
\end{equation}
as are the scattered far-field patterns described by vectors~$\M{f}_n = t_n \M{a}_n$ and either of them can be made real-valued~\cite{Gustafsson_etal2021_CMAT_Part1}. This can advantageously be used to expand pressure fields in driven problems. To that point, assume that an acoustic scatterer was driven by a field described by arbitrary vector~$\M{a}$. The scattered field is described by~\eqref{eq:fta}.
Expanding vector~$\M{f}$ into characteristic vectors~$\M{f}_n$, using orthogonality of eigenvectors and the symmetry of transition matrix~$\M{T}$ immediately leads to
\begin{equation}
    \M{f} = \sum \limits_n t_n \dfrac{\M{f}_n^\T{T} \M{a}}{\M{f}_n^\T{T} \M{f}_n} \M{f}_n . 
    \label{eq:fExp}
\end{equation}

This expansion shows that only characteristic modes with high absolute value~$|t_n|$ may contribute significantly to the scattered pressure, leading to~$|t_n|$ being referred to as modal significance. Since in typical scenarios, see Sec.~\ref{sec:Examples}, the modal significances~$|t_n|$ can be ordered into a rapidly converging series, only a few characteristic modes commonly contribute to the scattering. The projections~$\M{f}_n^\T{T} \M{a}$ then show how to excite these significant modes. 

\section{Limiting cases of hard and pressure-release boundaries}

The preceding formulation is applicable to arbitrary lossless surface impedances $Z_\T{s}$, but it reduces to special forms in the case of pressure release (soft, $\hat{y} \to \pm \J \infty$) or zero-velocity (hard, $\hat{y} = 0$) boundaries.  Because a scattering problem depends only on the contrast between the background medium and the obstacle, the two extreme cases of perfect hard and pressure release boundaries represent dual problems involving high and low acoustic impedance materials.  Considering the impedances of air ($413$ Rayl) and water ($1.5\times10^6$ Rayl) under normal conditions, these dual problems represent an approximation of two important scattering scenarios: air bubbles in water (pressure release boundary) and water droplets in air (hard boundary).

\subsubsection{Hard boundary ($\hat{y} = 0)$}
Hard boundaries characterized by vanishing normal velocity, e.g., a water droplet in air, are described by \eqref{eq:general-bem} with vanishing surface admittance $\hat{y} = 0$.  This limit simplifies the transition matrix to
\begin{equation}
        \M{T} = \J k_0 Z_0 \M{U}^p \M{D}^{-1} \M{U}^{\T{d}p,\T{H}}
        \label{eq:t-bem-hard}
\end{equation}
and the matrices used in the characteristic mode eigenvalue problem to 
\begin{equation}
    \begin{aligned}
        \M{X} &= \T{Im} \left\{ \J \M{D} \right\}, \\
         \M{R} &= \T{Re} \left\{ \J \M{D} \right\}.
    \end{aligned}
\end{equation}
Because the matrix $\M{D}$ is not symmetric, the characteristic surface pressure vectors $\M{p}_n$ are not, in general, orthogonal in the matrices $\M{X}$ or $\M{R}$ and cannot be made equiphase.  Nevertheless, the expression in \eqref{eq:t-bem-hard} represents an explicit, simple link between T-matrix and BEM numerical techniques.



\subsubsection{Pressure-release boundary ($\hat{y} \to \pm \J \infty$)}
For a pressure-release boundary, such as an air bubble in water, the normalized surface admittance tends toward infinity, and the boundary element equation in~\eqref{eq:general-bemGalerkin} should be rewritten using \eqref{eq:surfImpedCond} in terms of the normal velocity
\begin{equation}
    \M{v} =  - \dfrac{\hat{y}}{Z_0} \M{p} 
\end{equation}
as
\begin{equation}
    \J k_0 Z_0 \M{S} \M{v} = -\M{p}^\T{inc}.
    \label{eq:BEM-pr}
\end{equation}
Under the limit~$\hat{y} \to \pm \J \infty$, the transition matrix reduces to
\begin{equation}
        \M{T} = \J k_0 Z_0  \M{U}^{\T{d}p}  \M{S}^{-1}\M{U}^{\T{d}p,\T{H}} 
\end{equation}
while the BEM form of the characteristic mode eigenvalue problem can be rewritten in terms of modal velocities
\begin{equation}
    \M{v}_n =  - \M{S}^{-1}\M{U}^{\T{d}p,\T{H}} \M{a}_n
\end{equation}
as
\begin{equation}
    \M{X}\M{v}_n = - \lambda_n \M{R}\M{v}_n
    \label{eq:softXR}
\end{equation}
with
\begin{equation}
\begin{aligned}
    \M{X} &= \T{Im}\{\T{i}\M{S}\},\\
    \M{R} &= \T{Re}\{\T{i}\M{S}\}.
    \end{aligned}
    \label{eq:orth-pr}
\end{equation}

These simplifications allow for an alternative expansion formula to \eqref{eq:fExp} in terms of BEM incident pressure~$\M{p}^\T{inc}$ and modal velocities $\M{v}_n$. To that point, assume an expansion of the driven velocity $\M{v}$ in terms of the modal velocities $\M{v}_n$
\begin{equation}
    \M{v} = \sum_n \alpha_n \M{v}_n.
    \label{eq:uExpansion}
\end{equation}
Substituting~\eqref{eq:uExpansion} into~\eqref{eq:BEM-pr}, left multiplication with~$\M{v}_m^\T{H}$ and use of orthogonality~$\M{a}_m^\T{H} \M{T} \M{a}_n = t_n \M{a}_m^\T{H} \M{a}_n = t_n \M{a}_n^\T{H} \M{a}_n \delta_{mn}$ leads to
\begin{equation}
     \alpha_n  =  \dfrac{ \M{v}_n^\T{H} \M{p}^\T{inc}}{ t_n^* \M{a}_n^\T{H}  \M{a}_n} =  \dfrac{- \M{v}_n^\T{H} \M{p}^\T{inc}}{ \left( 1 - \J \lambda_n \right)  k_0 Z_0 \M{v}_n^\T{H} \M{R} \M{v}_n},
\end{equation}
where~\eqref{eq:softXR} was also used. Normalizing characteristic velocities as
\begin{equation}
    k_0 Z_0 \M{v}_n^\T{H}\M{R} \M{v}_n = 1,
\end{equation}
(a common choice in electromagnetics when the electric field integral equation is used to study perfectly conducting scatterers~\cite{harrington1971theory}), the expansion formula simplifies to
\begin{equation}
\M{v} = \sum_n \dfrac{-\M{v}_n^\T{H}\M{p}^\T{inc}}{1-\T{i}\lambda_n} \M{v}_n.
\label{eq:softFinalExpansion}
\end{equation}
For this case, characteristic modes diagonalize both the transition matrix (inherent for all lossless surface admittances, see \eqref{eq:a-orth}) as well as the system matrix $\M{S}$ governing the BEM problem itself.

Analogously to~\eqref{eq:fExp},  the expansion coefficient $\alpha_n$ in~\eqref{eq:softFinalExpansion} scales with the modal significance $|t_n|$, with a maximum (assuming stationary amplitudes of characteristic velocities) at resonance ($\lambda_n = 0$).  Also, the modal excitation coefficient depends directly on the alignment represented by the inner product of the modal velocity and the incident pressure field.  These two features are commonly used in electromagnetics to identify and tune modes to resonance and prescribe specific incident field distributions to excite particular characteristic modes~\cite{li2022synthesis}.

\section{Examples}
\label{sec:Examples}

In this section, we present three examples to demonstrate different aspects of characteristic mode decomposition via BEM and T-matrix methods.  The first example, a spherical shell, serves primarily to validate both formulations against an analytical benchmark while introducing common formats for interpreting characteristic mode data.  This is followed by two examples involving more general geometries: an open, multi-mode rigid tube and a single-mode Helmholtz resonator.

\subsection{Spherical obstacle}
\label{sec:sphere}

For spherically symmetric structures, scalar spherical waves~\eqref{eq:swExp} diagonalize the acoustic differential and integral operators and are, therefore, eigenstates of matrix~$\M{T}$ (characteristic modes) as well as of the BEM operators. For the case of a single spherical admittance shell of radius $a$, the characteristic eigenvectors~$\M{a}_n$ are independent of the surface admittance $\hat{y}$, though the eigenvalues~$t_n$ are specific to particular choices of the admittance
\begin{equation}
t_n = - \dfrac{\T{j}'_n(k_0a) + \J\hat{y}\T{j}_n(k_0a)}{\T{h}^{(1)\prime}_n(k_0a) + \J \hat{y}\T{h}_n^{(1)}(k_0a)},\quad n = 0,~1,~2,...
\label{eq:tn-sphere}
\end{equation}
where~$\T{j}_n $ denote spherical Bessel functions, while~$\T{h}_n^{(1)}$ denote outgoing spherical Hankel functions.
The special cases of hard and pressure release boundaries reduce to 
\begin{equation}
\begin{aligned}
    t_n(\hat{y}=0) &= -\dfrac{\T{j}'_n(k_0a)}{\T{h}^{(1)\prime}_n(k_0a)}, \\
   t_n(\hat{y}\to \pm \J\infty) &= -\dfrac{\T{j}_n(k_0a)}{\T{h}^{(1)}_n(k_0a)},
   \label{eq:te-tm-sphere}
   \end{aligned}
\end{equation}
with $2 n + 1$ multiplicity of degenerated modes.  Although recurrence relations~\cite[Eq. (10.51.1)]{NIST:DLMF} lead to a similarity in one set of eigenvalues across both problems $t_0(\hat{y}=0) = t_1(\hat{y}\to \pm \J\infty)$, the eigenvectors (either characteristic pressure fields or characteristic velocity fields) associated with these modes correspond to different order spherical harmonics with differing shapes.

Comparing the eigenvalues~\eqref{eq:tn-sphere} and \eqref{eq:te-tm-sphere} with the same spherical arrangement in electromagnetics, we realize that the pressure release boundary has identical spectrum of eigenvalues as transverse electric modes~\cite{CapekEtAl_ValidatingCMsolvers}, the only difference being the lack of an~$n=0$ ``monopole'' mode, which does not exist for (vectorial) electromagnetic fields. The \mbox{0-th} order mode also differs significantly in its asymptotic behavior in small wave size~$k_0a \to 0$. For~$n > 0$, the eigenvalues $t_n$ scale equally in acoustics and electromagnetics as~$\propto \pm \J (k_0 a)^{2n + 1}$, with the positive sign belonging to acoustically hard objects (or transverse magnetic modes) and with negative sign belonging to pressure release boundary (or transverse electric modes). For~$n = 0$, however, the hard material boundary breaks this rule, and its eigenvalue scales as~$\propto - \J (k_0 a)^3$, changing the sign and also decreasing two orders faster than that of the pressure release boundary. Since modal significance measures the strength of the scattering response, this result shows that, for small wave sizes~$k_0a \to 0$, the pressure release boundary should scatter significantly more than the hard boundary when excited by the 0-th order mode. This is reminiscent of the strong scattering of air bubbles in water, see~\cite[Chap.~8.2]{MorseTheoreticalAcoustics} though here we consider only the bubble's effective surface impedance and neglect nonlinear effects, such as deformation of air bubbles due to incident waves, see~\cite{Leighton1994} for more detailed discussion.

\begin{figure}
    \centering
    \includegraphics[width=3.25in]{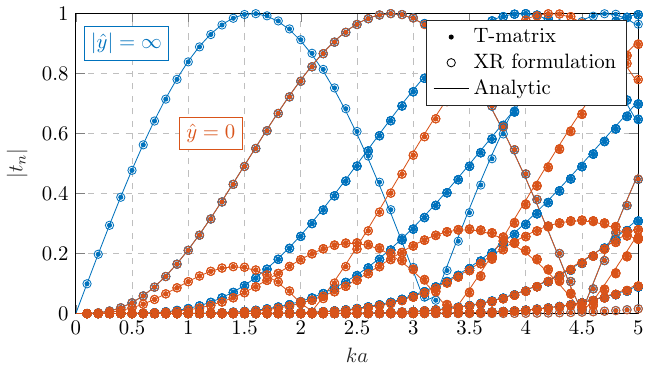}\\
    \includegraphics[width=3.25in]{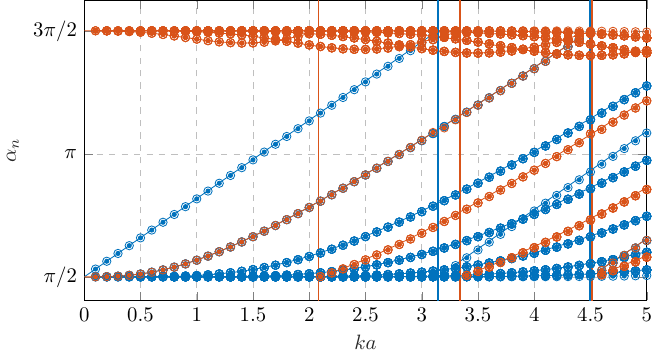}
    \caption{Modal significances $|t_n|$ (top) and characteristic angles $\alpha_n$ (bottom) for a sphere with soft (blue) and hard (red) boundary conditions calculated by the scattering and integral formulations.  Analytic results are shown as a benchmark.  Discontinuities in characteristic angles (zeros in modal significance) are denoted with solid vertical lines in the lower panel.}
    \label{fig:ms-sphere}
\end{figure}

\begin{figure}
    \centering
    \includegraphics[width=3.5in]{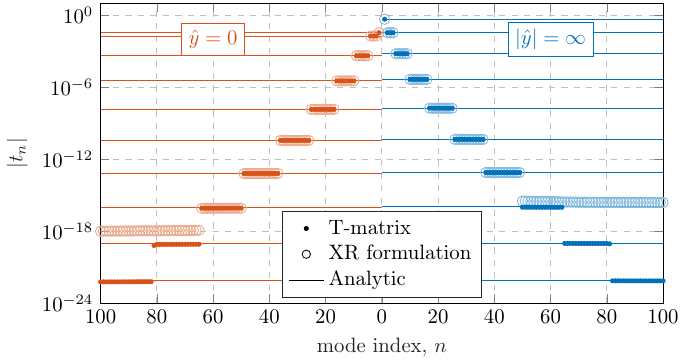}
    \caption{Dynamic range of modal significances for a sphere with soft (blue) and hard (red) boundary conditions at $ka = 0.5$.  Analytic results are shown as horizontal lines as a benchmark.}
    \label{fig:dr-sphere}
\end{figure}

To verify the formulations in Sec.~\ref{sec:cm-bem}, the characteristic modes of a spherical impedance shell are calculated using the analytic result in~\eqref{eq:tn-sphere},  the $\M{T}$ matrix formulation in~\eqref{eq:t-eig}, and the integral operator formulation in \eqref{eq:xr} for the two extreme cases of hard ($\hat{y}=0$) and pressure release ($\hat{y}\to\pm \J \infty$) boundaries.  A basis of 864 constant functions $\{\psi_n\}$, each defined over individual triangular domains within a mesh approximating the spherical suface, is used to compute all integral operators.  The triangle vertices are generated using Lebedev quadrature of degree 434~\cite{LebedevLaikov_QuadratureRuleSphere}. The resulting modal significances and characteristic angles are shown in Fig.~\ref{fig:ms-sphere}.  Good agreement is observed, validating the algebraic equivalence of the two numerical formulations and analytical benchmark.  This is further studied in Fig.~\ref{fig:dr-sphere}, which shows close agreement between the two methods and their analytic benchmark over a broad dynamic range at a fixed wave size.  
While the two methods are algebraically equivalent, the integral formulation exhibits poorer dynamic range due to the conditioning of the matrix $\M{R}$, associated with scattering and itself having eigenvalues accumulating at zero~\cite{tayli2018accurate,capek2022computational}.

\subsection{Open tube}
\begin{figure}
    \centering
    \includegraphics[width=\linewidth]{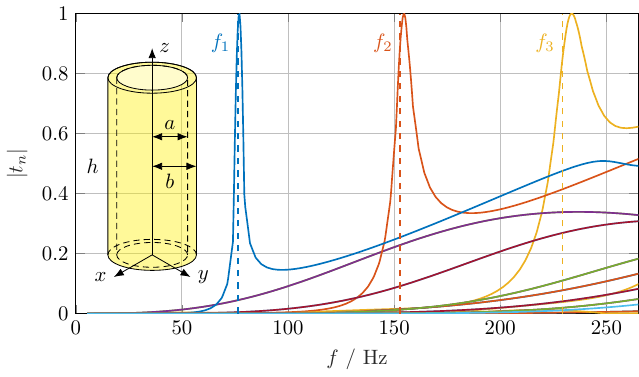}
    \caption{Modal significance of a rigid hollow tube with dimensions $h = 2\,\T{m}$, $a = 0.2\,\T{m}$, and $b = 0.25\,\T{m}$.  Dashed lines indicate analytic approximations to the structure's resonant frequencies given by \eqref{eq:tube-res-approx}.}
    \label{fig:tube-eigs}
\end{figure}

\begin{figure*}[ht]
    \centering
    \begin{tikzpicture}
        \node[align=center] at (0,0) {$f = 77$~Hz\\ \includegraphics[height=1.75in]{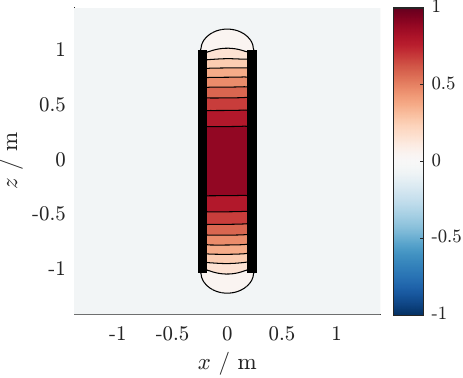}};
        \node[align=center] at (6,0) {$f = 154$~Hz\\\includegraphics[height=1.75in]{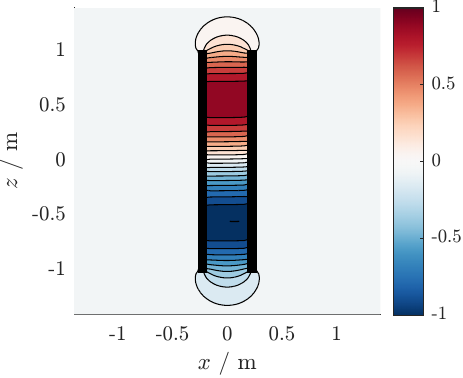}};
        \node[align=center] at (12,0) {$f = 233$~Hz\\  \includegraphics[height=1.75in]{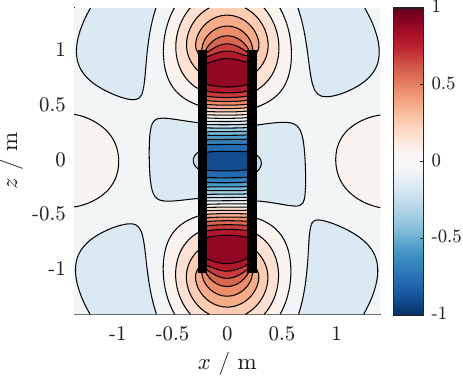}};
    \end{tikzpicture}
    \caption{Dominant characteristic modes of a hard open tube, presented as snapshots of normalized modal total pressures $\T{Re}\{p^\T{tot}_n\}$ in the $xz$ plane, near their modal resonance frequencies. Cross-sections of the tube walls are denoted by a black shaded regions.
}
    \label{fig:tube-ptot}
\end{figure*}

As a more general example, we consider a hard ($\hat{y}=0$) cylinder of length $h$, inner radius $a$, and outer radius $b$ in air ($c_0 = 343$~m/s, $\rho_0 = 1.225$~kg/m$^3$), see inset in Fig.~\ref{fig:tube-eigs}. The structure is centered at the origin, aligned with the $z$ axis, and discretized using 1800 triangles, each pertaining to a unique constant basis function.  Characteristic modes of the structure are plotted in Fig.~\ref{fig:tube-eigs} as a function of the excitation frequency, where the narrow-band resonance properties of several modes are shown within the studied band.  The resonant frequencies of these modes are compared to analytic approximation of resonance frequencies of natural modes
\begin{equation}
    f_n \approx \dfrac{nc_0}{2(h + 1.22 a)}
    \label{eq:tube-res-approx}
\end{equation}
which correspond to the tube length being integer multiples of half-wavelengths with an acoustic length correction due to non-zero radiation impedance \cite{Levine1948}. Note that the tube radiates at both ends. Hence, the length correction term is twice the usual value of $0.61a$. The detunings of the modal significances in Fig. \ref{fig:tube-eigs} from the theoretical predictions given by \eqref{eq:tube-res-approx} are due to the use of the simplest length correction model, which assumes infinitely thin walls and a limit of very low frequencies, see the discussion in~\cite{Ando1969,Bernard1996}, and due to the loose connection of characteristic modes and natural modes. The coincidence of frequencies of natural modes and characteristic modes is not general. Natural modes, being eigenstates of the Helmholtz operator, yield modal field distributions associated with eigenvalues representing complex-valued resonant frequencies. In contrast, characteristic modes are evaluated using an eigenvalue problem parameterized by a real frequency, which can be swept (as in this example) to locate real-valued resonant frequencies where the characteristic numbers $\lambda_n$ cross through zero.

At modal resonance, the scattering by the object is strong, as seen from expansion~\eqref{eq:fExp}. The bandwidth of this strong response can be evaluated from the characteristic data at the resonance frequency by expanding the modal significance in its Taylor series near its resonance frequency~$\omega_0$ and keeping only the two leading terms
\begin{equation}
    \left| t_n \right| = \dfrac{1}{ \sqrt {1 + \lambda_n^2 }} \approx 1 - \dfrac{1}{2} Q_n^2 \left( \dfrac{\omega  - \omega_0}{\omega_0} \right)^2,
\end{equation}
where the Q-factor~\cite{Harrington1972}
\begin{equation}
    Q_n = \omega_0 \left. \dfrac{ \partial \lambda_n}{\partial \omega } \right|_{\omega_0} = \omega_0 \left. \dfrac{ \partial \alpha_n}{ \partial \omega } \right|_{\omega_0}
\end{equation}
quantifies the relative bandwidth of the resonant peak and, therefore, of the scattering response. Hence, the scattering bandwidth can be estimated from an eigenvalue derivative at a single frequency, which is, in general, much less computationally expensive than a broadband frequency sweep.  

From the characteristic excitations $\M{a}_n$, the total characteristic pressures can be computed as a sum of incident and scattered contributions.  A snapshot of the instantaneous characteristic pressure fields $\T{Re}\{p^\T{tot}_n\}$ is shown in Fig.~\ref{fig:tube-ptot} for the three resonant modes in Fig.~\ref{fig:tube-eigs} at their corresponding resonant frequencies.  For all three modes, the large internal scattered field amplitude dominates the incident contribution. Furthermore, the large internal fields correspond to the predicted trend of standing waves at integer multiples of half-wavelengths.


Note that the total pressure on the surface of the structure corresponds to the modal surface pressures represented by the eigenvectors $\M{p}_n$.  Near resonance, all of these pressure distributions exhibit relatively high magnitudes on the interior of the tube relative to the exterior surface of the tube.  The contrast between internal and external pressure magnitudes in Fig.~\ref{fig:tube-ptot} roughly corresponds to the modal Q-factor of each mode at resonance since the internal fields are predominantly associated with reactive internal energy, while the external fields correspond to re-radiated energy.

\begin{figure}
    \centering
    \includegraphics[width=\linewidth]{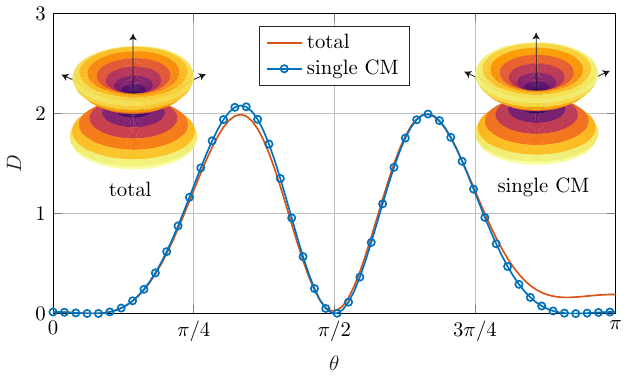}
    \caption{Far-field plots (three-dimensional plot and a two-dimensional cut along the spherical~$\theta$ direction) of the scattered pressure resulting from the incidence of a plane wave on the hollow tube from Fig.~\ref{fig:tube-eigs}. For comparison, the modal pressure corresponding to the dominant characteristic mode is also shown.}
    \label{fig:tube-expansion}
\end{figure}

As a last study related to this example, we present the fast convergence of modal expansion~\eqref{eq:fExp} when operating near resonance. To that point, assume that the tube is excited by a pressure plane wave along the $z$-axis at a frequency corresponding to the second resonance peak in Fig.~\ref{fig:tube-eigs}. The resulting directivity index $D$~\cite[\S 7.6]{KinslerFundamentalsOfAcoustics} and scattered pressure are shown in Fig.~\ref{fig:tube-expansion}. The result of the expansion~\eqref{eq:fExp}, with only the dominant modal contribution taken (a single term of the summation), is shown in the same figure. The two patterns agree for nearly all angles, with peak discrepancy in the backward direction due to the asymmetry of the total field compared to the symmetric characteristic field.  These results demonstrate how characteristic modes can be used as a sparse basis for predicting resonant scattering from arbitrarily shaped obstacles. 

\subsection{Helmoltz resonator}

The last example presents a geometrically more involved scatterer, a stiff multi-neck Helmholtz resonator~\cite{Langfeldt2019} in air, which exhibits a narrow band resonance at a small wave size. This is a scenario in which characteristic mode analysis excels in predicting the scattering behavior. 

\begin{figure}
    \centering
    \includegraphics[width=3.25in]{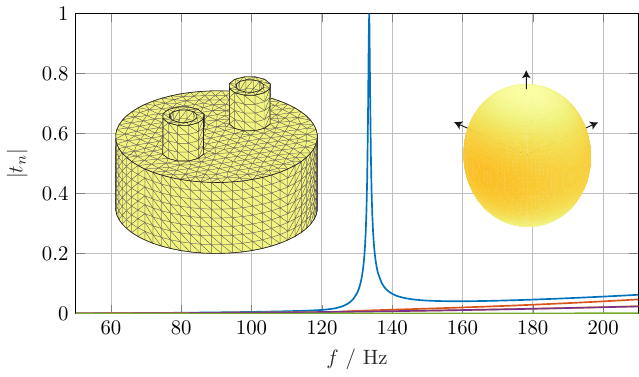}
    \caption{Modal significances $|t_n|$ calculated for a double-neck Helmholtz resonator. The cavity is made of a cylinder with radius~$R = 165 \,\T{mm}$ and height~$h = 0.7R$. The openings are identical and positioned symmetrically with respect to the cavity. The length of each opening is~$L = 0.5R$, and the radius of the opening is~$R_\T{o} = 0.15R$. The resonator is made of ideally hard material with wall thickness~$t = 0.075 R$.}
    \label{fig:ms-Helmholtz}
\end{figure}

\begin{figure}
    \centering
    \includegraphics[width=3.25in]{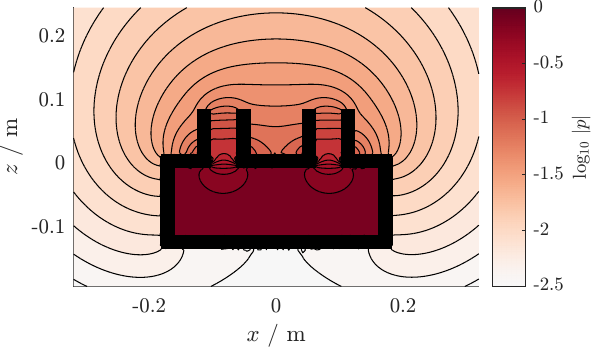}
    \caption{Normalized magnitude (logarithmic scale) of the dominant modal pressure distribution of a Helmholtz resonator at its resonance near 133 Hz.}
    \label{fig:helmholtzPressure}
\end{figure}

The modal significances are plotted in~Fig.~\ref{fig:ms-Helmholtz} revealing the expected resonance. The inset also presents the three-dimensional plot of the far-field pressure corresponding to the dominant mode at its resonance, showing mostly monopolar radiation. The dominant characteristic mode also well predicts the internal pressure of the resonator, see Fig.~\ref{fig:helmholtzPressure}, with almost constant pressure in the main cavity. 

\begin{figure}
    \centering
    \includegraphics[width=3.25in]{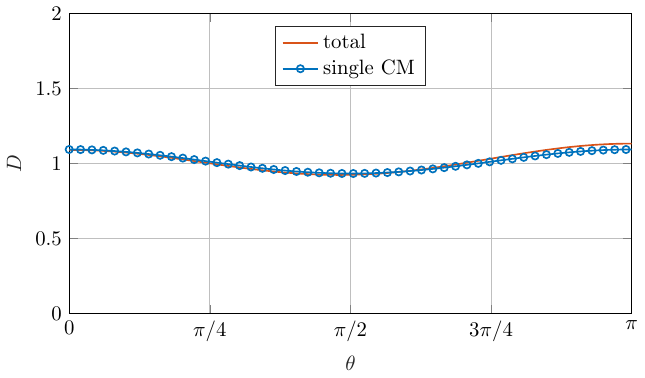}
    \caption{Comparison of the far-field plots of the scattered pressure resulting from the incidence of a plane wave on the Helmholtz resonator and from the dominant characteristic mode.}
    \label{fig:helmholtzCompar}
\end{figure}

When a characteristic mode resonates at such a low wave size, the modal significances of the non-dominant modes are low. The expansion formula~\eqref{eq:fExp} then predicts that the pressure field of the dominant mode approximates the scattering well even when excited by arbitrary excitation with nonzero projection into the dominant mode. As an example, Fig.~\ref{fig:helmholtzCompar} shows the comparison of the modal scattered pressure and scattered pressure caused by a plane wave propagating along the resonator axis.  Both the total and modal fields are essentially monopole patterns, with minimal variation over all scattered directions.  The total and single mode representations of the scattered fields do not deviate more than by 4~\%, with peak error occurring in the backward direction.

Knowing that the device operates in a single-mode regime provides additional opportunities to utilize characteristic mode analysis. For example, precisely finding the resonant frequency typically requires repetitive system matrix inversion. Alternatively, we can look for the characteristic angle~$\alpha_n$ passing~$\pi$, which is the resonance condition for characteristic mode. The characteristic angle can easily be interpolated with only a few frequency points~\cite{Gustafsson_etal2021_CMAT_Part2}. In connection with the implicitly restarted Arnoldi method to evaluate the dominant characteristic number, this offers considerable speed-up.

\section{Other Integral Formulations}

The definition of characteristic modes via transition matrix~\eqref{eq:t-eig} is independent of the numerical method used to analyze the object under test. That is, any arbitrary differential or integral acoustic solver can be used to assemble the transition matrix of a scatterer.

The integral formulations are, however, special in that they generally follow the same scheme as shown in Sec.~\ref{sec:CMfromBEM}. For example, under the assumption of constant mass density, a scattering from an  inhomogeneous distribution of  compressibility~$\kappa \left( \V{r} \right)$ can be described by an integral equation
\begin{equation}
    - \dfrac{ m \left( \V{r} \right)}{\rho_0 \left( \kappa \left( \V{r} \right)  - \kappa_0 \right)} + \omega^2 \int\limits_V  m \left( \V{r}' \right) g\left( \V{r} - \V{r}' \right) \T{d} V'  = p^\T{inc} \left( \V{r} \right),
    \label{eq:volIE}
\end{equation}
where~$\kappa_0$ is the background compressibility and~\mbox{$m \left( \V{r} \right) =  - \rho _0 \left( \kappa \left( \V{r} \right)  - \kappa _0 \right) p \left( \V{r} \right)$} are the equivalent mass sources representing the scatterer~\cite{MorseTheoreticalAcoustics}. Using Galerkin's method and expansion
\begin{equation}
    m \left( \V{r} \right) = \sum \limits_n M_n \psi_n \left( \V{r} \right),
\end{equation}
the integral equation~\eqref{eq:volIE} is rewritten as
\begin{equation}
    \left( \M{\Psi } + \omega^2 \M{S} \right) \M{M} =  - \M{p}^\T{inc},
\end{equation}
where vector~$\M{M}$ collects expansion coefficients~$M_n$, matrix
\begin{equation}
    \Psi_{mn} = \int \limits_V \dfrac{ \psi_m \left( \V{r} \right) \psi_n \left( \V{r} \right)}{\rho _0 \left( \kappa \left( \V{r} \right) - \kappa_0 \right)} \T{d} V
\end{equation}
represents the material distribution and matrices~$\M{S}, \M{p}^\T{inc}$ are the same as in the previous sections only with volumetric integration.

Abbreviating~$\M{Z} = \J \left( \M{\Psi } + \omega^2 \M{S} \right)$, following the same steps as in Sec.~\ref{sec:cm-bem} and~\ref{sec:CMfromBEM} and switching from surface integration to volume integration, it is then straightforward to show that transition matrix is evaluated as
\begin{equation}
    \M{T} = - k_0 Z_0 \omega^2 \M{U}^{\T{d}p} \M{Z}^{-1} \M{U}^{\T{d}p,\T{H}}.
\end{equation}
Furthermore, in lossless scenarios where mass density and compressibility are real-valued, characteristic modes can also be obtained from a generalized eigenvalue problem
\begin{equation}
    \M{X} \M{M}_n = -\lambda_n \M{R} \M{M}_n 
\end{equation}
with~$\M{R} = \T{Re}\left\{ \M{Z} \right\}$ and $\M{X} = \T{Im} \left\{ \M{Z} \right\}$. Since the system matrix~$\M{Z}$ is symmetric in this specific formulation and, therefore, $ \M{M}_m^\T{T} \M{R} \M{M}_n = \M{M}_n^\T{T} \M{R} \M{M}_n \delta_{mn}$, then apart from general expansion~\eqref{eq:fExp}, one can also directly expand the unknown vector~$\M{M}$ into modal characteristic contributions as
\begin{equation}
    \M{M} =  \sum\limits_n \dfrac{- \J \M{M}_n^\T{T} \M{p}^\T{inc}}{ \left( 1 - \J \lambda _n \right) \M{M}_n^\T{T} \M{R} \M{M}_n} \M{M}_n.
\end{equation}

\section{Conclusion}

Characteristic mode decomposition of acoustic scatterers yields several convenient properties, which, though differing slightly in interpretation, carry many benefits that have made characteristic modes a popular tool in electromagnetic analysis and design.  In particular, they diagonalize the BEM impedance operator, yield orthogonal scattered fields, and allow for the estimate of scattering bandwidths from derivatives of modal quantities at single frequencies.  

Additionally, the theory of characteristic modes is a convenient link between boundary element methods and scattering (T-matrix) approaches to numerical analysis.  The algebraic connections between these techniques allow not only for the computation of characteristic modes for arbitrary objects using BEM, but also suggest straightforward methods for obtaining the T-matrix of complex physical systems via BEM.  This broadens the applicability of both analysis techniques by leveraging their complementary strengths for varying problem types.

\begin{acknowledgments}
This work was supported by the Czech Science Foundation under project~\mbox{No.~24-11678S}. The work of Viktor Hruska was supported by the Czech Science Foundation under project~\mbox{No.~22-33896S}.
\end{acknowledgments}

\appendix

\section{Explicit Relations of the Used Boundary Element Method}
\label{app:A}

For any closed surface not containing sources of sound, the scattered pressure can be evaluated as an addition of single and double-layer terms~\cite{MorseTheoreticalAcoustics}
\begin{equation}
\begin{aligned}
    p^\T{scatt}_p \left\{ p \right\} \left( \V{r} \right) &= \oint\limits_{\partial V} p \left( \V{r}' \right) \V{n} \left( \V{r}' \right) \cdot \nabla' g \left( \V{r} - \V{r}' \right) \T{d} S' \\
    p^\T{scatt}_{\T{d}p} \left\{ \V{n} \cdot \nabla p  \right\} \left( \V{r} \right) &= -\oint\limits_{\partial V} \V{n} \left( \V{r}' \right) \cdot \nabla' p \left( \V{r}' \right) g\left( \V{r} - \V{r}' \right) \T{d} S',
    \label{eq:scattOper}
\end{aligned}
\end{equation}
where~$g \left( \V{r} - \V{r}' \right)$ is background Green's function.

The total pressure just outside the surface~$\partial V$ is therefore subject to the following condition~\cite{MorseTheoreticalAcoustics}
\begin{equation}
    p \left( \V{r} \right) - p^\T{scatt}_p \left\{ p \right\} \left( \V{r} \right) - p^\T{scatt}_{\T{d}p} \left\{ \V{n} \cdot \nabla p  \right\} \left( \V{r} \right) = p^\T{inc} \left( \V{r} \right),
\end{equation}
from which the integral equation~\eqref{eq:general-bem} results.

Galerkin's procedure~\cite{Kantorovich1982,KirkupTheBoundaryElementMethodInAcoustics} with basis functions~$\psi_n$ then results in matrix system~\eqref{eq:general-bemGalerkin}, with
\begin{equation}
    \begin{aligned}
    D_{m n} &= - \int\limits_{S_m} \psi_m \psi_n\T{d}S + \int\limits_{S_m} \psi_m p^\T{scatt}_p \left\{ \psi_n \right\} \T{d}S \\
    S_{m n} &= \int\limits_{S_m} \psi_m p^\T{scatt}_{\T{d}p} \left\{ \psi_n \right\} \T{d}S \\
    p_{m}^{\T{inc}} &= \int\limits_{S_m} \psi_m p^\T{inc}\T{d}S,
    \end{aligned}
\end{equation}
where~$S_n$ denotes the support of the basis function~$\psi_m$.  In all numerical examples in this paper, we implement the above expressions using constant basis functions defined over individual mesh triangles.

When free-space Green's function is factorized into  spherical waves~\cite[Sec.~9.6]{Jackson_ClassicalElectrodynamics}, the scattering operators~\eqref{eq:scattOper} are factorized in the same way which results in operators 
\begin{equation}
    \begin{aligned}
    U_{\alpha n}^p &= - \dfrac{\J}{\sqrt{Z_0}} \int\limits_{S_n} \psi_n \left( \V{r}' \right)\V{n}\left( \V{r}' \right) \cdot \nabla' \left(\T{u}_\alpha^{\left( 1 \right)} \left( \V{r}' \right)\right)^\ast \T{d}S' \\
    U_{\alpha n}^{\T{d}p} &= \dfrac{\J}{\sqrt{Z_0}} \int\limits_{S_n} \psi_n \left( \V{r}' \right) \left(\T{u}_\alpha^{\left( 1 \right)} \left( \V{r}' \right) \right)^\ast  \T{d} S' 
    \end{aligned}
\end{equation}
transforming surface pressure into expansion coefficients of outgoing spherical waves.

\section{Power Balance}
\label{app:B}

The net cycle mean power passing along the outer normal of a surface~$S$ circumscribing the scatterer can be evaluated as
\begin{equation}
    \dfrac{1}{2} \oint\limits_S \T{Re} \left\{ \V{v} p^* \right\} \cdot \T{d} \V{S} = - P^\T{lost}
\end{equation}
and equals to the negative of the cycle mean power lost in the scatterer~\cite{MorseTheoreticalAcoustics}. Assuming that the integration surface lies in a space free of sources, the velocity field~$\V{v}$ can be evaluated as
\begin{equation}
    \V{v} = - \dfrac{\J}{ k_0 Z_0} \nabla p
\end{equation}
and the lost power can be evaluated as
\begin{equation}
   P^\T{lost} = - \dfrac{1}{2 k_0 Z_0} \oint\limits_S \T{Im} \left\{  p^* \nabla p \right\} \cdot \T{d} \V{S}.
\end{equation}
Assuming spherical expansion~\eqref{eq:swExp} and integrating over a spherical surface gives~$P^\T{lost} = P^\T{ext} - P^\T{scatt}$, where
\begin{equation}
    P^\T{ext} = - \dfrac{1}{2} \sum\limits_\alpha  \T{Re} \left\{ a_\alpha^* f_\alpha  \right\} 
\end{equation}
is the extinct cycle mean power and
\begin{equation}
    P^\T{scatt} = \dfrac{1}{2} \sum\limits_\alpha  \left| f_\alpha \right|^2
\end{equation}
is the cycle mean scattered power.

Employing relation~\eqref{eq:fta}, the lost power can be written as
\begin{equation}
    P^\T{lost} = - \dfrac{1}{2} \T{Re} \left\{ \M{a}^\T{H} \left( \M{T} + \M{T}^\T{H} \M{T} \right) \M{a}  \right\}.
\end{equation}
Since the above relation must be valid for all vectors~$\M{a}$, it can be seen that for a lossless scenario with~$P^\T{lost} = 0$, matrix~$\M{T}$ must be a normal matrix satisfying~$\M{T} \M{T}^\T{H} = \M{T}^\T{H} \M{T}$.

\bibliography{references,acousticsX}

\end{document}